# Properties of Sparse Distributed Representations and their Application to Hierarchical Temporal Memory

Subutai Ahmad and Jeff Hawkins
Numenta
Redwood City, CA 94063
sahmad@numenta.com, jhawkins@numenta.com

March 24, 2015

**Abstract**

Empirical evidence demonstrates that every region of the neocortex represents information using sparse activity patterns. This paper examines Sparse Distributed Representations (SDRs), the primary information representation strategy in Hierarchical Temporal Memory (HTM) systems and the neocortex. We derive a number of properties that are core to scaling, robustness, and generalization. We use the theory to provide practical guidelines and illustrate the power of SDRs as the basis of HTM. Our goal is to help create a unified mathematical and practical framework for SDRs as it relates to cortical function.

## I. Introduction

Our neocortex continually processes an endless stream of rich sensory information. It does this remarkably well, better than any existing computer. A wealth of empirical evidence demonstrates that cortical regions represent all information using sparse patterns of activity. To function effectively throughout a lifetime these representations must have tremendous capacity and must be extremely tolerant to noise. However, a detailed theoretical understanding of the capacity and robustness of cortical sparse representations has been missing.

Cortical representations are such that only a small percentage of neurons are very active while the rest remain relatively inactive, and hence the activity is considered *sparse*; the result of inhibitory neurons. The representations are considered *distributed* because the information is encoded not just in a single neuron but across a set of active neurons. Sparse distributed representations (SDRs) encode information throughout the cortex, and for a diverse variety of functions. The representations exist in early auditory and visual areas (Hromádka et al., 2008; Weliky et al., 2003), corresponding to sensory features such as audio frequencies, and visual lines and edges. The SDRs in later sensory areas encode more abstract and categorical information. Similarly, both concrete motor commands (Sanes & Donoghue, 2000) and abstract behavioral planning (Graziano et al., 2002) are represented via sparse distributed encodings in primary motor and premotor areas, respectively.

This paper analyzes certain properties and operations of SDRs relevant to cortical encoding. The binary SDR operations examined here are motivated by neuroscience and central to generalization, high order sequence memory, and achieving invariant representations in HTM learning algorithms. We derive classification error bounds and capacity scaling laws based on these operations. The results demonstrate that neurons encoding information via SDRs have massive representational power. Robustness to noise is high enough such that reliable classification can be performed with as much as 50% noise. We analyze the "union property" and show how it can be used to make multiple temporal predictions reliably.

The theory behind sparse representations has become a topic of significant interest in recent years. They have been studied in the context of theoretical neuroscience (Kanerva, 1988, Olshausen and Field, 2004), statistics (Tibshirani, 1996), machine learning (Lee, 2008), compressed sensing (Donoho, 2006), and even web server data structures (Broder and Mitzenmacher, 2004). In our analysis we lean on the intuitions provided by Kanerva (Kanerva, 1988, Kanerva, 1997) as well as some of the techniques used for analyzing Bloom filters (Bloom, 1970). This discussion follows in Section II. The application of these properties to a number of different aspects of the HTM Learning Algorithm, as described in (Hawkins and Ahmad, 2010), is discussed in Section III. Section IV concludes the paper with the advantages of SDRs as the mathematical base of HTM.

## II. Mathematical Foundations of SDRs

In this section we discuss certain mathematical properties of SDRs, with a focus on deriving fundamental scaling laws and error bounds. Properties such as probability of mismatches, robustness in noise, subsampling, classifying vectors, and unions demonstrate the usefulness of SDRs as a memory space. We illustrate properties and operations with examples to give an intuition to the mathematics.

### A. Definitions and Notation

*SDRs:* Given a population of $n$ neurons, their instantaneous activity is represented as an SDR, i.e. an $n$-dimensional vector of binary components, e.g. $x = [b_0, \ldots, b_{n-1}]$. Typically these vectors are highly sparse ,i.e. a small percentage of the components are 1. We use $w_x$ to denote the number of components in $x$ that are 1, i.e. $w_x = \|x\|_1$.

*Overlap*: We determine the similarity between two SDR encodings using an overlap score. The overlap score is simply the number of bits that are ON in the same locations in both vectors. If $x$ and $y$ are two binary SDRs, then the overlap can be computed as the dot product:

$$overlap(x, y) \equiv x \cdot y$$

Notice we do not use a typical distance metric, such as Hamming or Euclidean, to quantify similarity. With overlap we can derive some useful properties discussed later, which would not hold with these distance metrics.

*Matching*: We realize a match between two SDRs if their overlap exceeds some threshold $\theta$:

$$match(x, y) \equiv overlap(x, y) \geq \theta$$

Typically $\theta$ is set such that $\theta \leq w_x$ and $\theta \leq w_y$.

Consider an example of two SDR vectors:

$$x = [0100000000000000000100000000000110000000]$$
$$y = [1000000000000000000100000000000110000000]$$

Both vectors have size $n = 40$, and $w = 4$. The overlap between $x$ and $y$ is 3, and thus the two vectors match when $\theta = 3$. Throughout the paper we'll present examples of SDRs with sample parameters typical of current Hierarchical Temporal Memory (HTM) implementations. These are listed below, and detailed later in Section III.

*HTM parameters*:
$n = 1024$ to $65{,}536$, representing the length of an SDR vector
$w = 10$ to $40$, representing the number of ON bits in an SDR vector
$s = 0.05\%$ to $2.0\%$, representing the sparsity, where $s = \frac{w}{n}$

## B. Uniqueness and Exact Matches

Given a fixed $n$ and $w$, the number of unique SDR encodings is $n$ choose $w$:

$$\binom{n}{w} = \frac{n!}{w!\,(n-w)!} \tag{1}$$

Note this is significantly smaller than the number of encodings possible with dense representations, which is $2^n$. This implies a potential loss of information, as the number of possible input patterns is much greater than the number of possible representations in the SDR encoding. In practice this is meaningless. With $n = 40$ and $w = 4$, the number of encodings is 91,390. With more typical values $n = 2048$ and $w = 40$, the SDR representation space is astronomically large at $2.37 \times 10^{84}$ encodings; the estimated number of atoms in the observable universe is $\sim 10^{80}$.

Given two random SDR encodings with the same parameters, $x$ and $y$, the probability they are identical is

$$P(x = y) = 1/\binom{n}{w} \tag{2}$$

With $n = 1024$ and $w = 2$ there are 523,776 possible encodings and the probability two random encodings are identical is rather high, i.e. 1 in 523,776. This probability decreases extremely rapidly as $w$ increases. With $w = 4$, the probability dives to less than 1 in 45 billion. For the HTM values $n = 2048$ and $w = 40$, the probability two random encodings are identical is essentially zero. These calculations are carried out for varying parameters in Table 1 of Appendix A.

## C. Overlap Sets

We introduce the notion of an *overlap set* to help analyze the effects of matching under varying conditions. Let $x$ be an SDR encoding. The overlap set of $x$ with respect to $b$ is $\Omega_x(n, w, b)$, defined as the set of vectors of size $n$ with $w$ bits on, that have exactly $b$ bits of overlap with $x$. The number of such vectors is $|\Omega_x(n, w, b)|$, the cardinality of the set. Assuming $b \leq w_x$ and $b \leq w$,

$$|\Omega_x(n, w, b)| = \binom{w_x}{b} \times \binom{n - w_x}{w - b} \qquad (3)$$

The first term in the product of (3) is the number of subsets of $x$ with $b$ bits ON, and the second term is the number of other patterns containing $n - w_x$ bits, of which $w - b$ bits are on.

## D. Inexact Matching

In general we would like the system to be somewhat tolerant to changes or noise in the input. That is, it is rare to require exact matches where $\theta = w$. Lowering $\theta$ decreases the sensitivity and increases the overall noise robustness of the system. For example, consider SDR vectors $x$ and $x'$, where $x'$ is corrupted by random noise. With $w = 40$ and $\theta$ lowered to 20, the noise can change 50% of the ON bits and still match $x$ to $x'$.

Yet increasing the robustness comes with the cost of more false positives. That is, decreasing $\theta$ also increases the probability of a false match with another random vector. There is an inherent tradeoff in these parameters, as we would like the chance of a false match to be as low as possible while retaining robustness. With appropriate parameter values the SDRs can have a large amount of noise robustness with a very small chance of false positives.

Given an SDR encoding $x$ and another random vector $y$, what is the probability of a false match, i.e. the chance the $overlap(x, y) \geq \theta$? A match is defined as an overlap of $\theta$ bits or greater, up to $w$. With $\binom{n}{w}$ total patterns, the probability of a false positive is:

$$fp_w^n(\theta) = \frac{\sum_{b=\theta}^{w} |\Omega_x(n, w, b)|}{\binom{n}{w}} \qquad (4)$$

Note an exact match occurs when $\theta = w$. The numerator in (4) evaluates to 1, and the equation reduces to (2).

For example, again suppose vector parameters $n = 1024$ and $w = 4$. If the threshold is $\theta = 2$, corresponding to 50% noise, then the probability of an error is one in 14,587. That is, with 50% noise there is a significant chance of false matches. If $w$ and $\theta$ are increased to 20 and 10, respectively, the probability of a false match decreases drastically to less than 1 in $10^{13}$. Thus, with a relatively modest increase in $w$ and $\theta$ (holding $n$ fixed), SDRs can achieve essentially perfect robustness with up to 50% noise. These calculations are carried out for varying parameters in Table 2 of Appendix A.

For the majority of cases we are interested in[1], the first term in the numerator sum of (4) dominates by at least an order of magnitude, thus (5) gives an excellent approximation of the false positive likelihood:

---

[1] That is, whenever $w > 7$ and $\theta > \frac{w}{2}$.

$$fp_w^n(\theta) \approx \frac{|\Omega_x(n, w, \theta)|}{\binom{n}{w}} \qquad (5)$$

**E. Subsampling**

An interesting property of SDRs is the ability to reliably compare against a subsampled version of a vector. That is, recognizing a large distributed pattern by matching a small subset of the active bits in the large pattern. Let $x$ be an SDR vector and let $x'$ be a subsampled version of $x$, such that $w_{x'} \leq w_x$. The subsampled vector $x'$ will always match $x$, as long as $\theta \leq w_{x'}$, but as you increase the subsampling the chance of a false positive increases.

What is the probability of a false match between $x'$ a random vector $y$? Here the overlap set is computed with respect to the subsample $x'$, rather than the full vector $x$. If $b \leq w_{x'}$ and $w_{x'} \leq w_y$ then the number of patterns with exactly b bits of overlap with $x'$ is:

$$|\Omega_{x'}(n, w_y, b)| = \binom{w_{x'}}{b} \times \binom{n - w_{x'}}{w_y - b} \qquad (6)$$

Given a threshold $\theta \leq w_{x'}$, the chance of a false positive then is:

$$fp_{w_y}^n(\theta) = \frac{\sum_{b=\theta}^{w_{x'}} |\Omega_{x'}(n, w_y, b)|}{\binom{n}{w_y}} \qquad (7)$$

Notice (6) and (7) differ from (3) and (4), respectively, only in the vectors being compared. That is, subsampling is simply a variant of the inexact matching properties discussed above.

For instance, suppose $n = 1024$ and $w_y = 8$. Subsampling half the bits in $x$ and setting the threshold to 2 (i.e. $w_{x'} = 4, \theta = 2$), we find the probability of an error is one in 3,142. However, increasing $w_y$ to 20 and the relevant parameter ratios fixed (i.e. $w_{x'} = 10, \theta = 5$) the chance of a false positive drops to 1 in 2.5 million. Increasing $n$ to 2048, $w_y = 40$, $w_{x'} = 20$, and $\theta = 10$, more practical HTM parameter values, the probability of a false positive plummets to better than 1 in $10^{12}$. This is remarkable considering that the threshold is about 25% of the original number of bits. These calculations are carried out for varying parameters in Table 2 of Appendix A.

**F. Classifying a Set of Vectors**

We consider a form of classification similar to nearest neighbor classification. Let $X$ be a set of $M$ vectors, $X = \{x_1, \dots, x_m\}$, where each vector $x_i$ is an SDR. We assume that all vectors in $X$ are unique with respect to matching, i.e.

$$\forall_{x \in X} \forall_{y \in X, y \neq x} match(x, y) = false \tag{8}$$

Given any vector $y$ we classify it as belonging to this set as follows:

$$y \in X \equiv \exists_{x_i \in X} match(x_i, y) = true \tag{9}$$

How reliably can we classify a vector $x_i$ corrupted by up to $t$ bits of noise? Assuming $t \leq w - \theta$, there are no false negatives in this scheme, only false positives. Our question is then what is the probability the classification of a random vector $y$ is a false positive? Since all vectors in $X$ are unique with respect to matching, the probability of a false positive is bounded by:

$$fp_x(t) \leq \sum_{i=1}^{M} fp_{w_{x_i}}^n(t) \tag{10}$$

This is an upper bound because it "double counts" a vector that matches two different vectors in $X$. In the special case where all vectors in X have identical $w$, this is just:

$$fp_x(\theta) \leq M fp_w^n(\theta) \tag{11}$$

Consider for example $n = 64$ and $w = 3$ for all vectors. If $\theta = 2$, 10 vectors can be stored in the list and the probability of false positives is about 1 in 22. Increasing $w$ and $\theta$ to 12 and 8, respectively, maintaining the ratio $\frac{\theta}{w} = \frac{2}{3}$, the chance of a false positive drops to about 1 in 2363. Now increase the parameters to more realistic values: $n = 1024$, $w = 21$, and $\theta = 14$ (i.e. two-thirds of $w$). In this case the chance of a false positive with 10 vectors plummets to about 1 in $10^{20}$. In fact, with these parameters the false positive rate for storing a billion vectors is better than 1 in $10^{12}$.

This result illustrates a remarkable property of SDRs. Suppose a large set of patterns is encoded in SDRs, and stored in a list. A massive number of these patterns can be retrieved almost perfectly, even in the presence of a large amount of noise. The main requirement being the SDR parameters $n$, $w$, and $t$ need to be sufficiently high. As illustrated in the above example, low values such as $n = 64$, $w = 3$, etc. are insufficient to capture these properties. These calculations are carried out for a variety of parameters in Table 3 of Appendix A.

## G. The Surprising Union Property

One of the most fascinating properties of SDRs is the ability to reliably store a set of patterns in a single fixed representation by taking the OR of all the vectors. We call this the "union property". To store a set of $M$ vectors, the union mechanism is simply the Boolean OR of all the vectors, resulting in a new vector $X$. To determine if a new SDR $y$ is a member of the set, we simply compute the $match(X, y)$.

$$x_1 = [01000000000010000000 \ldots 010]$$
$$x_2 = [00000000000000000010 \ldots 100]$$
$$x_3 = [10100000000000000000 \ldots 010]$$
$$\vdots$$
$$x_{10} = [00000000000000110000 \ldots 010]$$

$$X = x_1 OR x_2 OR, \ldots, x_{10}$$

$$X = [11100000000110110000 \ldots 110]$$

$$y = [10000000000001000000 \ldots 001]$$

$$\therefore match(X, y) = 1$$

Figure 1: (top) Taking the OR of a set of $M$ SDR vectors results in the union vector $X$. With the sparsity of each individual vector at 2%, and $M = 10$, it follows that the sparsity of $X$ is at most 20%. The logic is straightforward: if there is no overlap within the set of vectors, each ON bit will correspond to its own ON bit in the union vector, summing the sparsities. With overlap, however, ON bits will be shared in the union vector, resulting in a lower sparsity. (bottom) Computing the $match(X, y)$ reveals if $y$ is a member of the union set $X$ – i.e. if the ON positions in $y$ are ON in $X$ as well.

The advantage of the union property is a fixed-size SDR vector can store a dynamic set of elements. As such, a fixed set of cells and connections can operate on a dynamic list. It also provides an alternate way to do classification. In HTM's, unions are used extensively to make temporal predictions, for temporal pooling, to represent invariances, and to create an effective hierarchy. However, there are limits on the number of vectors that can be reliably stored in a set. That is, the union property has the downside of increased potential for false positives.

How reliable is the union property? There is no risk of false negatives; if a given vector is in the set, its bits will all be one regardless of the other patterns, and the overlap will be perfect. However, the union property increases the likelihood of false positives. With the number of vectors, $M$, sufficiently large, the union set will become saturated with ON bits, and almost any other random vector will return a false positive match.

Let us first calculate the probability of a false positive assuming exact matches, i.e. $\theta = w$. In this case, a false positive with a new random pattern $y$ occurs if all of the bits in $y$ overlap with $X$. When $M = 1$, the probability any given bit is 0 is given by $1 - \frac{w}{n}$. As $M$ grows, this probability is given by

$$p_0 = \left(1 - \frac{w}{n}\right)^M \qquad (12)$$

After $M$ union operations, the probability a given bit in X is ON is $1 - p_0$.

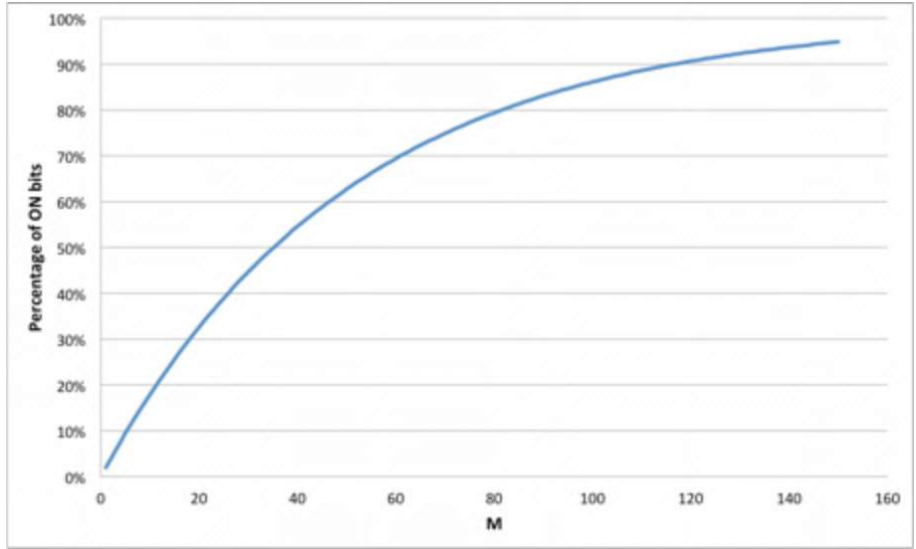

Figure 2: The expected percentage of ON bits as a function of $M$, the number of vectors in the union set. In this case, $N = 2048$ and $w = 40$, i.e. $s = 2\%$.

The probability of a false positive, i.e. all $w$ bits in $y$ are ON, is therefore

$$p_{fp} = (1 - p_0)^w = \left(1 - \left(\frac{w}{n}\right)^M\right)^w \tag{13}$$

The technique used to arrive at (13) is similar to the derivation of the false positive rate for Bloom filters (Bloom, 1970; Broder and Mitzenmacher, 2004)[2].

For instance, consider SDR parameters $n = 1024$ and $w = 2$. Storing $M = 20$ vectors, the chance of a false positive is about 1 in 680. However, if $w$ is increased to 20, the chance drops dramatically to about 1 in 5.5 billion. This is a remarkable feature of the union property. In fact, if increasing $M$ to 40, the chance of an error is still better than $10^{-5}$.

To gain an intuitive sense of the union property, the expected number of ON bits in the union vector is $n(1 - p_0)$. This grows slower than linearly, as shown in Figure 2. Consider for instance $M = 80$, where 20% of the bits are 0. When we consider an additional vector with 40 ON bits, there is a reasonable chance it will have at least one bit among this 20, and hence it won't be a false positive. That is, only vectors with all of their $w$ bits amongst the 80% ON are false positives. As we increase $n$ and $w$, the number of patterns that can OR together reliably increases substantially. Calculations to illustrate this property are carried out for a variety of parameters in Table 4 of Appendix A.

---

[2] The slight difference being in Bloom filters each bit is chosen independently, i.e. with replacement. As such, a given vector could contain less than $w$ ON bits. In this analysis we guarantee that there are exactly $w$ bits on in each vector.

In HTMs the vector sizes are often substantially larger than 1024, and the union property proves incredibly useful. We will discuss the impact of these numbers in more detail in Section III.

**H. Inexact Matches with Unions**

The union property was derived above in the context of no noise, i.e. $\theta = w$. We would like to understand the effects of noise in the system, and therefore now relax the criteria of a perfect match when considering unions. How reliably then can we match $y$ against $X$ using a threshold $\theta < w$?

As mentioned above, the expected number of ON bits in the union vector $X$ is $\widetilde{w}_X = n(1 - p_0)$. Assuming $n \geq \widetilde{w}_X \geq w$, we can calculate the expected size of the overlap set:

$$E[|\Omega_X(n,w,b)|] = \binom{\widetilde{w}_X}{b} \times \binom{n - \widetilde{w}_X}{w - b} \tag{14}$$

In order for a match, we can have an overlap of $\theta$ or greater bits (up to $w$). The probability of a false match is therefore:

$$\varepsilon \approx \frac{\sum_{b=\theta}^{w} |\Omega_X(n,w,b)|}{\binom{n}{w}} \tag{15}$$

Note (15) is an approximation of the error, as we're working with the expected number of ON bits in $X$[3]. As in (4), for most of the cases we are interested in, the first term in the numerator sum dominates, allowing us to simplify (15). Thus, (16) is an excellent approximation for all practical purposes[4].

$$\varepsilon \approx \frac{|\Omega_X(n,w,b)|}{\binom{n}{w}} \tag{16}$$

As one would expect, the chance of error increases as the threshold is lowered. The consequences of this tradeoff can be mitigated by increasing $n$. Suppose $n = 1024$ and $w = 20$. When storing $M = 20$ vectors, the chance of a false positive when using perfect matches is about 1 in 5 billion. Using a threshold of 19 increases the false positive rate to about 1 in 123 million. When $\theta = 18$, the chance increases to 1 in 4 million. However if you increase $n$ to 2048 with $\theta = 18$, the false positive rate improves dramatically to 1 in 223 billion. This example illustrates the union property's robustness to noise, and is yet another example of our larger theme: small linear changes in SDR numbers can cause super-exponential improvements in the error rates.

---

[3] Subsequently, in the simplified case of no unions, i.e. $w_x = w$, there are slight discrepancies with (14), which is an exact calculation. Nevertheless, for numbers in the ranges we are concerned with, the discrepancies are insignificant. An excellent approximation of the error for all practical purposes is given by (16).

[4] Specifically whenever $w > 12$ and $\theta > \frac{w}{2}$.

### I. Computational Efficiency

Although SDR vectors are large, all the operations we've discussed run in time linear in the number of ON bits. That is they are $O(w)$ and independent of the size of the vector, $n$. This would not be the case, however, with more standard distance metrics which are typically $O(n)$. For HTM systems this is important since in practice $w \ll n$.

## III. SDRs and HTM

Hierarchical Temporal Memory is a detailed computational theory of the neocortex. At the core of HTM are time-based learning algorithms that store and recall spatial and temporal patterns. SDR is the primary data structure used in the cortex, and used everywhere in HTM systems. The algorithms as described in the HTM whitepaper (Hawkins and Ahmad, 2010) are critically reliant on binary SDRs. In HTM there isn't a single unitary SDR, rather there are a number of distinct SDRs for specialized purposes. These SDRs are involved in different parts of the HTM model neuron, and different functions in the system of algorithms.

The following discussion assumes familiarity with HTM basics. It would behoove the reader to understand the implementation and pseudocode in the HTM whitepaper (Hawkins and Ahmad, 2010), as we discuss SDRs in the context of the neuron model i.e. columns of cells, and their proximal and distal dendritic segments, and cortical operations – i.e. spatial pooling and predictions.

Note that HTM structures and operations such as hierarchy, feedback, sensorimotor inference, and motor commands also use specific SDRs and rely on SDR properties. Although a discussion of those topics as well as the learning algorithm is beyond the scope of this paper, the same principles and intuitions apply.

### A. Notation

The main HTM algorithms discussed are "Spatial Pooling" (SP) and "Temporal Memory" (TM), where TM refers to the sequence learning portion of the pseudocode in Chapter 4 of HTM whitepaper (Hawkins and Ahmad, 2010) The following constants are used:

$N$ : size of input vector
$C$ : number of columns
$k$ : the number of columns active after spatial inhibition
$L$ : number of cells per column
$S$ : number of segments per cell

### B. Spatial Pooling

Here we provide an overview of the SDR operations as they are implemented in the SP process, with vector operations depicted in Figure 3. The SP process takes as input a binary vector of length $N$. In practice this vector is usually sparse, but is not required to be sparse. Each column in the SP represents proximal segments in a cell. A binary matrix with dimensions $N \times C$ can be used to represent the set of connected synapses in the SP. This matrix is not restricted to be sparse, but in practice is almost always sparse. The result of the vector-matrix multiplication is a vector of overlap counts, the third data frame in Figure 3. An inhibition step chooses the column winners to form the output. That is, the indices corresponding to the top $k$ overlaps correspond to the ON bits in the output SDR. By construction this SDR is of size $1 \times C$ with $k$ components that are 1.

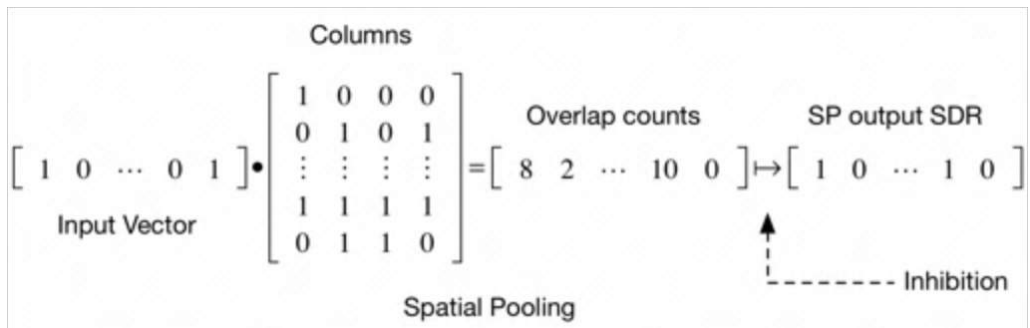

Figure 3: SDRs used in spatial pooling, where SP inhibition is run on the input vector to yield an output binary vector representing the columns with the top overlap counts.

In the absence of learning, the SP process examines the overlap between a set of randomly initialized columns and individual binary input vectors. The top $k$ columns, determined by calculating the overlap, win and form the ON bits in the resulting SDR. This SDR is then used as input into subsequent TM processes.

Consider a collection of random binary vectors $X$ each with size $n$ and $w_x$ ON bits. Given a new random sparse vector $y$ of the same size, but possibly different $w$, how many vectors in $X$ will have exactly $b$ bits of overlap with $y$? The probability of such a match for a single vector in $X$ is:

$$p(overlap(x,y) = b) = \frac{|\Omega_y(n, w_x, b)|}{\binom{n}{w_x}} \qquad (17)$$

It follows that the expected number of columns with $b$ bits of overlap is $|X| \cdot p(overlap(x,y) = b)$, from which we can calculate the "overlap curve". This is the overlap for each column after sorting, shown in Figure 4 shows this curve for a representative set of parameters. This curve provides some intuition as to how a randomly initialized SP process functions. One can see that a small amount of noise at the input will somewhat change the output SDR, but that most of the winners will be unchanged. The sharper the drop off after k, the more noise robustness in the system. One of the effects of SP learning is to make the overlap curve sharper, thereby increasing robustness to noise.

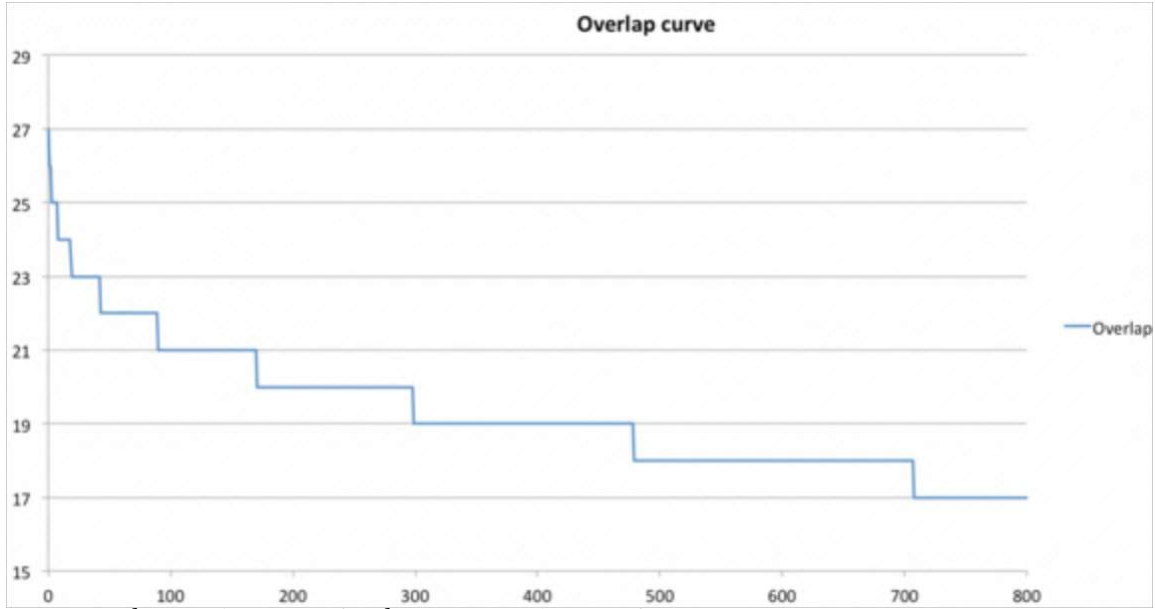
Figure 4: The overlap curve for the top 800 columns of 2048.

## C. Temporal Memory

The TM process is discussed in two phases, where the first phase calculates the active states at time step $t$, and the second predicts which states will be active in the next time step, $t + 1$.

The first phase of TM follows the SP and determines the current active temporal state. This phase examines the cells in each column predicted in the previous time step, and the actual winning columns in the SP output SDR. Cells in winning columns that were correctly predicted stay ON to represent the active temporal state of the system. Figure 5 shows a schematic of this operation. The first matrix is the $1 \times C$ SP output SDR, followed by the $L \times C$ matrix predicted state matrix. Row-by-row element-wise multiplication (represented by the symbol $\odot$) of these two matrices is used to determine the active state, which is also $L \times C$. All three structures are sparse.

$$\begin{bmatrix} 0 & 1 & \cdots & 0 & 1 \end{bmatrix} \odot \begin{bmatrix} 1 & 0 & 0 & 0 \\ 0 & 1 & 0 & 1 \\ \vdots & \vdots & \vdots & \vdots \\ 1 & 1 & 0 & 1 \\ 0 & 1 & 1 & 0 \end{bmatrix} = \begin{bmatrix} 0 & 0 & 0 & 0 \\ 0 & 1 & 0 & 1 \\ \vdots & \vdots & \vdots & \vdots \\ 0 & 1 & 0 & 1 \\ 0 & 1 & 0 & 0 \end{bmatrix}$$

SP Output SDR     Predicted State     Active State

Temporal Memory Phase 1

Figure 5: SDRs used in phase 1 of temporal memory, where the inputs are the spatial-pooled SDR and the predicted state, and the output is the active state.

The second phase of TM determines the predictions for the next time step by matching individual cell segments with the current active state. Figure 6 shows this operation schematically. Each cell contains a list of segments, where each segment is represented by an SDR of size $L \times C$. For each segment the TM computes a match between this representation and the current active state. If the match of any segment is above the threshold, the predicted state for this cell will be ON.

$$\text{Segment List}$$

$$\begin{bmatrix} 1 & 0 & \cdots & 0 \end{bmatrix} \bullet \begin{bmatrix} \begin{bmatrix} 1 & 0 & \cdots & 0 \\ 0 & 1 & \cdots & 0 \\ 0 & 0 & \cdots & 1 \\ 1 & 0 & \cdots & 1 \end{bmatrix} \\ \vdots \\ \begin{bmatrix} 0 & 0 & \cdots & 1 \\ 1 & 1 & \cdots & 0 \\ 0 & 0 & \cdots & 1 \\ 1 & 1 & \cdots & 0 \end{bmatrix} \end{bmatrix} = \begin{bmatrix} \begin{bmatrix} 8 & 0 & \cdots & 9 & 0 \end{bmatrix} > \tau \\ \vdots \\ \begin{bmatrix} 3 & 1 & \cdots & 4 & 0 \end{bmatrix} > \tau \end{bmatrix} \Rightarrow \begin{bmatrix} 1 & 0 & \cdots & 0 \end{bmatrix}$$

Active State   Predicted State

Temporal Memory Phase 2

Figure 6: SDRs in phase two of temporal memory, where the inputs are the active state and set of segments, and the output is the predicted state.

Matching the active state is equivalent to the SDR classification function discussed in the previous section. The ability of each cell to robustly "classify" the active state will impact the overall performance of the TM. In addition, at time $t$ there may be many possible states predicted for time $t + 1$, and all the cells corresponding to these states will be in predicted state. From a computational standpoint, this step then determines *a union of possible temporal states* for the next time step. The ability of the predicted state to robustly represent this union is a fundamental property of SDRs and key to the performance of the TM.

The TM learns sequences of SDRs, but what is the mechanism by which HTM uses SDRs to represent high-order sequences? The TM can uniquely represent inputs as unique steps in a sequence by using the cells within columns to represent inputs in varying contexts. That is, in a learned temporal sequence, only one of the cells in an HTM column becomes active. High-order sequences rely on creating a unique temporal context. We can calculate the number of unique temporal contexts based on the number of cells per column. For $w$ columns of $x$ cells each, there are $x^w$ ways to represent the same input in different contexts. For conservative HTM parameters $w = 40$ and $x = 8$, this evaluates to $1.33 \times 10^{36}$. This is a massive capacity, even with a limited number of cells per column; some HTM models implement 32 cells per column.

The full set of segments implement classification as discussed above in Section IIF. The TM is able to make temporal predictions if and only if enough cells are able to correctly classify the current active state. In TM each of a cell's dendrite segments connects to other cells in the region, but the connections are limited to only a small portion of the cells. A basic distal segment (without overloading or pooling) uses inexact matching, as discussed above. The equations (4)-(7) give us lower bounds on the number of synapses per segment and help us understand the effect of lowering the activation threshold. We can use the equations to set these parameters such that false classification by a segment is highly unlikely. With more segments per dendrite, the TM can learn more unique contexts. A higher activation threshold will call for more of the synapses on these segments to be active in order for a cell to become predicted, placing a more stringent requirement on segments' subsampling their input.

The TM makes a union of predictions. Using the union property we can figure out how many predictions can be made at a time with minimal probability of false matches. We can provide upper bounds on the number of simultaneous predictions before the predicted state saturation becomes useless – i.e. predict too many states.

# IV. Conclusion

To fully understand any computational system, we need an understanding of the data structure and their properties. SDRs form the basic data structure in HTM systems. From their mathematics we can derive properties such as bounds and scaling laws, performance characteristics, and ideal parameters. In this paper we have shown that SDRs can be used to perform robust classification under noise and random deletions. The union property shows how dynamic lists can be robustly represented within a single fixed vector. We have outlined how various HTM operations map to SDRs.

Under the right set of parameters SDRs enable a massive capacity to learn temporal sequences and form the basis for highly robust classification systems. Taken together these properties provide a simple and elegant mathematical theory characterizing the robustness of HTM systems.

# Acknowledgement

The authors thank Alexander Lavin and Celeste Baranski for their help in editing this paper, and Yuwei Cui, Scott Purdy, and Chetan Surpur for many helpful discussions.

# Appendix A – Practical Considerations

**The Numbers Game**

The equations derived above provide significant insights to set most of the parameters in HTM networks. They can tell us the number of columns, the desired sparsity, how to determine the potential pool percent, how to set the threshold and other parameters on a segment, the limits of temporal memory predictions, how many patterns we can reliably pool together, and so on.

Getting HTMs to work well in practice often boils down to a numbers game, and this fact is often lost. A common approach in working with HTMs is to start with toy problems and small values of *n, w*, etc. This suffices for getting the code working and initial debugging, but it is critical to move on to more practical values. As shown in the paper, there are a number of exponentials and super-exponentials involved in SDR properties. With a little bit of parameter tweaking, it's easy to find range of numbers that work well across a large number of applications.

In the rest of this Appendix, we list tables with a number of different parameter combinations and associated error rates. They can be used to understand the behavior and set parameters appropriately.

**SDR Actuarial Tables**

*Table 1 – Exact matches*: This table shows the number of possible patterns in various SDRs and the associated probability of false exact matches. Even small values of *n* and *w* lead to a large universe of patterns and low error rates.

| n | w | Number of patterns | Prob. of false match |
|---|---|---|---|
| 64 | 1 | 64 | 0.015625 |
| 64 | 3 | 41664 | 2.40015E-05 |
| 64 | 5 | 7624512 | 1.31156E-07 |
| 64 | 7 | 621216192 | 1.60975E-09 |
| 64 | 9 | 27540584512 | 3.631E-11 |
| 64 | 11 | 7.43596E+11 | 1.34482E-12 |
| 512 | 1 | 512 | 0.001953125 |
| 512 | 3 | 22238720 | 4.49666E-08 |
| 512 | 5 | 2.87516E+11 | 3.47807E-12 |
| 512 | 7 | 1.75619E+15 | 5.69416E-16 |
| 512 | 9 | 6.20812E+18 | 1.61079E-19 |
| 1024 | 1 | 1024 | 0.000976563 |
| 1024 | 3 | 178433024 | 5.60434E-09 |
| 1024 | 5 | 9.29119E+12 | 1.07629E-13 |
| 1024 | 7 | 2.29479E+17 | 4.35769E-18 |
| 1024 | 9 | 3.29326E+21 | 3.03651E-22 |

*Table 2 – Inexact matches*: This table shows the probability of false inexact matches when comparing two SDRs and various values of the threshold t. The values show it is difficult to get a low false match probability with n=64. However, increasing n to 1024 dramatically decreases the chances even with a much lower threshold. It is obvious larger values of n will continue to improve the chances, and are thus excluded from the table.

| n | w | t | Prob. of false match |
|---|---|---|---|
| 64 | 4 | 4 | 1.57387E-06 |
| 64 | 4 | 3 | 0.000379303 |
| 64 | 4 | 2 | 0.017093815 |
| 64 | 4 | 1 | 0.232525308 |
| 64 | 8 | 8 | 2.25929E-10 |
| 64 | 8 | 7 | 1.01442E-07 |
| 64 | 8 | 6 | 9.84351E-06 |
| 64 | 8 | 5 | 0.000360558 |
| 64 | 8 | 4 | 0.006169265 |
| 64 | 32 | 32 | 5.45666E-19 |
| 64 | 32 | 24 | 6.70223E-05 |
| 64 | 32 | 16 | 0.59857385 |
| 1024 | 20 | 20 | 1.82484E-42 |
| 1024 | 20 | 17 | 3.50023E-31 |
| 1024 | 20 | 14 | 9.93621E-23 |
| 1024 | 20 | 10 | 9.32924E-14 |

*Table 3 – Classifying vectors*: This table shows the probability of false matches when classifying a list of vectors under different parameter combinations. It's clear increasing the value of n and w has a very large impact on the number of vectors that can be classified accurately and robustly.

| n | w | M | $\theta$ | Prob. of false positive |
|---|---|---|---|---|
| 64 | 3 | 10 | 3 | 0.000240015 |
| 64 | 3 | 10 | 2 | 0.044162826 |
| 64 | 12 | 10 | 12 | 3.04487E-12 |
| 64 | 12 | 10 | 10 | 2.68378E-07 |
| 64 | 12 | 10 | 8 | 0.000423112 |
| 1024 | 21 | 10 | 21 | 3.81689E-43 |
| 1024 | 21 | 10 | 14 | 8.8349E-21 |
| 1024 | 21 | 1.00E+09 | 21 | 3.81689E-35 |
| 1024 | 21 | 1.00E+09 | 17 | 9.5841E-21 |
| 1024 | 21 | 1.00E+09 | 14 | 8.8349E-13 |

*Table 4 – Union with inexact matches*: This table shows the probability of false matches when you have a union of patterns and when the threshold is allowed to vary. We highlight some specific parameter settings corresponding to some common Temporal Memory regimes. n=8192 corresponds to 1024 columns with 8 cells per column. n=65536 corresponds to 2048 columns with 32 cells per column.

| n | w | t | M | Prob. of false match |
|---|---|---|---|---|
| 64 | 4 | 4 | 10 | 0.043131941 |
| 64 | 4 | 3 | 10 | 0.42652697 |
| 64 | 8 | 8 | 10 | 0.07104513 |
| 64 | 8 | 7 | 10 | 0.866750589 |
| 1024 | 20 | 20 | 20 | 1.2532E-10 |
| 1024 | 20 | 18 | 20 | 2.49E-07 |
| 1024 | 20 | 16 | 20 | 6.59353E-05 |
| 1024 | 20 | 20 | 30 | 7.76674E-08 |
| 1024 | 20 | 18 | 30 | 8.09221E-05 |
| 1024 | 20 | 16 | 30 | 0.011323733 |
| 1024*8=8192 | 20 | 20 | 60 | 4.33389E-18 |
| 1024*8=8192 | 20 | 18 | 60 | 4.61314E-14 |
| 1024*8=8192 | 20 | 16 | 60 | 6.58692E-11 |
| 1024*8=8192 | 20 | 14 | 60 | 2.95373E-08 |
| 1024*8=8192 | 40 | 40 | 80 | 2.15567E-20 |
| 1024*8=8192 | 40 | 36 | 80 | 1.92105E-13 |
| 1024*8=8192 | 40 | 32 | 80 | 1.57968E-08 |
| 2048*32=65536 | 40 | 40 | 80 | 1.06052E-53 |
| 2048*32=65536 | 40 | 36 | 80 | 1.9751E-43 |
| 2048*32=65536 | 40 | 32 | 80 | 3.37299E-35 |
| 2048*32=65536 | 40 | 28 | 80 | 4.95502E-28 |
| 2048*32=65536 | 40 | 40 | 1000 | 2.4446E-14 |
| 2048*32=65536 | 40 | 36 | 1000 | 5.40046E-08 |
| 2048*32=65536 | 40 | 32 | 1000 | 0.001114562 |
| 2048*32=65536 | 40 | 28 | 1000 | 1 |
| 2048*32=65536 | 40 | 40 | 600 | 2.86956E-21 |
| 2048*32=65536 | 40 | 36 | 600 | 3.07893E-14 |
| 2048*32=65536 | 40 | 32 | 600 | 3.06773E-09 |
| 2048*32=65536 | 40 | 28 | 600 | 2.67369E-05 |

With n=64 it is difficult to do any significant pooling; the error rate is simply too high. With the more typical settings of 1024 or 2048 columns, a very large amount of pooling is possible with very low error rates and reasonable noise robustness.